\documentclass[twocolumn, aps, prl, superscriptaddress, floatfix]{revtex4-2}

\usepackage[utf8]{inputenc}
\usepackage{amsmath, amssymb, bm}
\usepackage{graphicx}
\usepackage{hyperref}
\usepackage{color}

\usepackage{subfigure,dcolumn,jabbrv}
\usepackage{graphicx}
\usepackage{dcolumn}
\usepackage{bm}
\usepackage{CJK}
\usepackage{url}
\usepackage{color}
\usepackage{booktabs}
\usepackage{natbib}
\usepackage{hyperref}
\usepackage{cleveref}

\usepackage{hyperref}
\usepackage[bb=ams, cal=cm, scr=boondox, frak=euler]{mathalpha}

\renewcommand{\v}[1]{\textbf{#1}}
\renewcommand{\rm}[1]{\textrm{#1}}

\begin{document}

\title{Universal EOS--Radius Inverse Mappings Govern Precision-Dependent Inference of the Neutron Star Equation of State}

\author{Bao-An Li}
\email{Bao-An.Li@etamu.edu}
\affiliation{Department of Physics and Astronomy, East Texas A\&M University, Commerce, TX 75429-3011, USA}

\date{\today}

\begin{abstract}
Bayesian inference of the neutron star (NS) equation of state (EOS) generally assumes that improved observations primarily reduce posterior uncertainties while leaving inferred EOS parameters unchanged. Using mock measurements of the radius of a canonical $1.4\,M_\odot$ NS with identical central values but varying observational precisions, we show that the inferred posterior means of EOS parameters can shift systematically as the measurement uncertainty changes. We demonstrate that this behavior originates from previously unidentified nearly universal inverse mappings between the NS radius $R_{1.4}$ and empirical EOS parameters. Across a broad range of observational precisions, posterior samples collapse onto nearly unique functions. These mappings are largely independent of observational precision and define a low-dimensional EOS manifold underlying Bayesian inference. We show that the precision dependence of inferred EOS parameters arises from nonlinear filtering of the posterior radius distribution through these mappings. In the narrow-distribution limit this effect reduces to a Jensen-type correction proportional to the local curvature of the inverse mapping, while for presently realistic uncertainties the full nonlinear-filtering relation accurately reproduces the posterior means. Our results reveal a geometric origin of precision-dependent inference in NS EOS studies and provide a new framework for connecting astrophysical observations directly to microscopic nuclear many-body theories.
\end{abstract}

\maketitle

\textit{Introduction.-}
Determining the equation of state (EOS) of supradense neutron-rich matter remains one of the central goals of modern nuclear astrophysics. Increasingly precise measurements of neutron star (NS) masses, radii, tidal deformabilities, and moments of inertia are providing unprecedented opportunities to constrain the EOS. Among these observables, the radius $R_{1.4}$ of a canonical $1.4\,M_\odot$ NS is particularly important because it is primarily determined by the pressure of neutron-rich matter around twice the nuclear saturation density, $2\rho_0$, where $\rho_0$ denotes the saturation density of symmetric nuclear matter (SNM) \cite{LattimerPrakash}. Lattimer and Prakash showed that the pressure near $2\rho_0$ approximately obeys the scaling relation $\rm{Pre}(2\rho_0)\propto R_{1.4}^{4}$, revealing a strongly nonlinear connection between NS observables and the underlying EOS \cite{LattimerPrakash}. More recently, this scaling was derived analytically from the intrinsic properties of the dimensionless Tolman--Oppenheimer--Volkoff (TOV) equations \cite{Tolman1939,Oppenheimer1939} without invoking any specific EOS model \cite{Cai23,Cai25,Cai26}, demonstrating that the observed nonlinearity originates fundamentally from NS structure itself.

Future observations are expected to improve NS radius measurements from the current typical uncertainty of about $1$ km to the $\sim0.1$ km level through next-generation X-ray timing missions \cite{eXTP,Ath} and gravitational-wave detectors \cite{Sat,Evans,Chat,Pac,Band,Fin,Walker}. Such an order-of-magnitude improvement is motivated by several major scientific goals \cite{Astro20}, including distinguishing competing nuclear many-body theories, identifying possible phase transitions and twin-star configurations, and establishing direct connections between NS observables and the microscopic properties of supradense matter \cite{Li24PRD,Xavier1,Li26APJ,Xavier2}. It is generally assumed that improved observational precision simply narrows the posterior distributions of EOS parameters. We show that increasing observational precision does more than reduce statistical uncertainties: it systematically changes the inferred posterior means because Bayesian inference is governed by nonlinear inverse mappings between NS observables and EOS parameters. 

Throughout this work, ``universal" denotes mappings that are insensitive to the microscopic realization of the EOS once identical physical constraints are imposed. In this Letter, we demonstrate that the nonlinear inverse mapping effect is both systematic and nearly universal. Using Bayesian inference with mock measurements of a canonical NS radius having an identical central value but different uncertainties, we discover a set of nearly universal inverse mappings between $R_{1.4}$ and six empirical EOS parameters. These mappings are largely independent of observational precision and define a low-dimensional inferred manifold underlying Bayesian EOS inference. We show that the precision dependence of inferred EOS parameters originates from nonlinear averaging over the inverse mappings. In the narrow-distribution limit, the effect reduces to a Jensen-type correction proportional to the local curvature of the mapping \cite{Jen,Ber,Ebook,infobook}, while for realistic uncertainties in the presently available data, the full nonlinear-filtering relation accurately reproduces the posterior means. Our results reveal a geometric origin of nonlinear-filtering corrections in EOS inference and provide a new framework for connecting NS observations directly to microscopic nuclear theories. 

The mappings identified here reveal that precision-dependent, nonlinear-filtering corrections are not mere technical peculiarities of Bayesian NS analyses, but rather generic geometric consequences of non-linear parameter inference. Crucially, this framework provides a unified perspective on parameter extraction across nuclear physics and astrophysics. Similar nonlinear filtering effects and Jensen-type shifts are expected whenever bulk EOS properties are mapped from finite-precision observables, including measurements of neutron skins of heavy nuclei across various experimental probes, collective excitations in nuclear structure studies, and heavy-ion collision dynamics. While we define ``universality'' within the scope of identical hadronic constraints, this remarkable property is physically grounded in the fact that the radius $R_{1.4}$ of a canonical NS is predominantly determined by the pressure around $2\rho_0$. At these intermediate densities, exotic degrees of freedom like quark matter are unlikely to have emerged, rendering the inverse mappings robust against variations in high-density phase transitions that typically only manifest in more massive stars~\cite{Li26APJ}. 
Although mathematical universality cannot be established for arbitrary EOSs, the observed collapse of EOS-radius inverse mappings demonstrates a highly constrained low-dimensional manifold governing all realistic nucleonic EOSs.\\

\textit{Nonlinear Filtering and Jensen Correction in Bayesian EOS Inference.-} For a sufficiently narrow Gaussian distribution of an observable $R=R_0+\delta R$ with mean $R_0$ and standard deviation $\sigma_R$, the expectation value of any nonlinear mapping $h(R)$ satisfies the Jensen expansion \cite{Jen,Ber,Ebook,infobook}
\begin{equation}
\langle h\rangle(\sigma_R)
\simeq
h(R_0)
+\frac12 h''(R_0)\sigma_R^2,
\label{eq:jensen}
\end{equation}
where the second term is the leading nonlinear correction. Its sign is determined by the local curvature $h''(R_0)$: convex mappings increase the means, whereas concave mappings decrease them. As shown below, this expression represents the high-precision limit of the more general nonlinear-filtering relation governing Bayesian EOS inference from NS radius data.

As summarized in the End Matter, we employ a metamodel EOS characterized by six empirical parameters $
\bm{h}=(J_0,K_0,J_{\rm{sym}},K_{\rm{sym}},L,E_{\rm{sym,0}})$, where $K_0$ and $J_0$ characterize properties of SNM while $L$, $K_{\rm{sym}}$, $J_{\rm{sym}}$ and $E_{\rm{sym},0}\equiv E_{\rm{sym}}(\rho_0)$ describe the density dependence of nuclear symmetry energy $E_{\rm{sym}}(\rho)$. More specifically, the energy per nucleon in neutron-rich matter can be written as \cite{Bom}
\begin{equation}
E(\rho,\delta)=E_0(\rho)+E_{\rm{sym}}(\rho)\delta^2+\mathcal{O}(\delta^4),
\end{equation}
where $\rho$ is the baryon density and $\delta=(\rho_n-\rho_p)/\rho$ is the isospin asymmetry. The energy per nucleon $E_0(\rho)$ in SNM and symmetry energy $E_{\rm{sym}}(\rho)$ are parameterized as
\begin{eqnarray}
E_0(\rho) &=& E_0(\rho_0)
+\frac{K_0}{2}\eta^2
+\frac{J_0}{6}\eta^3,\\
E_{\rm{sym}}(\rho) &=&
E_{\rm{sym}}(\rho_0)
+L\eta
+\frac{K_{\rm{sym}}}{2}\eta^2
+\frac{J_{\rm{sym}}}{6}\eta^3,
\end{eqnarray}
with $E_0(\rho_0)=-16$ MeV and $\eta\equiv(\rho-\rho_0)/(3\rho_0)$. The six EOS parameters constitute the EOS parameter space explored in the present work. These parameters are varied over broad prior ranges consistent with current astrophysical, nuclear laboratory and theoretical constraints as discussed in detail in Refs. \cite{Li24PRD,Xavier1,Xavier2,Li26APJ}.

For charge neutral nucleons and electrons at $\beta$ equilibrium before muons appear, the pressure is given by \cite{Lat01}
\begin{equation}\label{pre}
P_{\rm{NS}}(\rho,\delta)
=\rho^2[\frac{dE_{\rm{0}}(\rho)}{d\rho}+\frac{dE_{\rm{sym}}(\rho)}{d\rho}\delta^2]
+\frac{1}{2}\delta(1-\delta)\rho E_{\rm{sym}}(\rho).
\end{equation}
The value of the isospin asymmetry $\delta$ (or the corresponding proton fraction $x_p=(1-\delta)/2$)
is determined completely by the symmetry energy via
\begin{eqnarray}\label{xp}
x_p(\rho)= 0.048 \left[E_{\rm{sym}}(\rho)/E_{\rm{sym}}(\rho_0)\right]^3
(\rho/\rho_0)(1-2x_p(\rho))^3.
\end{eqnarray}
The NS pressure $P_{\rm{NS}}(\rho,\delta)$ becomes barotropic, once the density profile of the proton fraction $x_p(\rho)$ is determined self-consistently by the symmetry energy $E_{\rm{sym}}(\rho)$. 

This framework provides a flexible and nearly model-independent representation of dense matter while preserving a transparent connection between NS observables and fundamental nuclear matter properties. In particular, the meta-model allows the posterior distributions of individual EOS parameters to be interpreted directly in terms of the underlying density dependence of SNM EOS and the symmetry energy, making it especially well-suited for investigating the nonlinear-filtering corrections in Bayesian inference of EOS parameters.

\begin{figure}[thb]
\vspace{-0.5cm}
\centering
 \resizebox{0.55\textwidth}{!}{
\includegraphics[width=1.\textwidth]{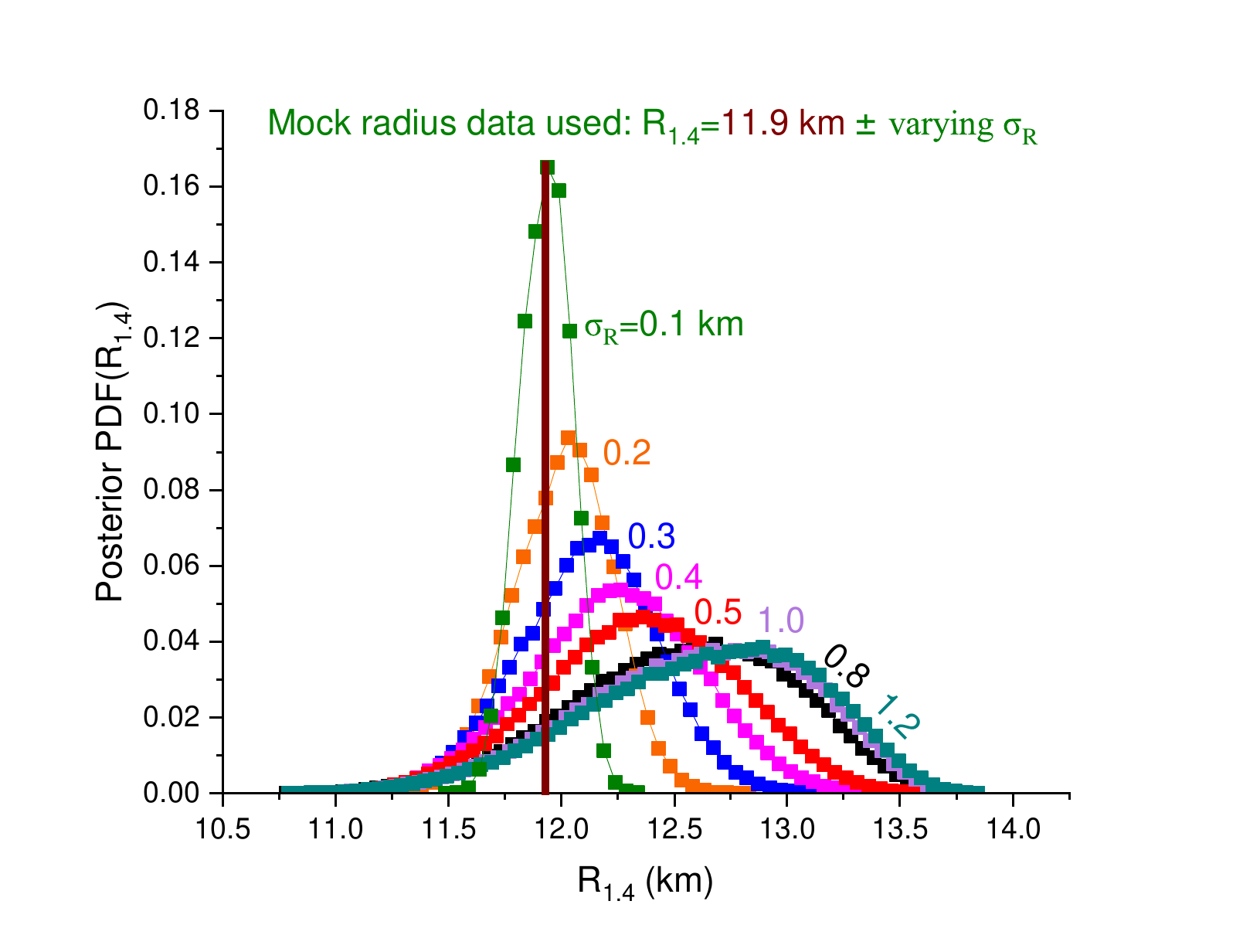}
}
\vspace{-1.cm}
\caption{Posterior probability distribution functions (PDFs) of $R_{1.4}$ inferred from the mock radius data $R_{1.4}$ having a Gaussian distribution with a mean value of $R_0=11.9~{\rm{km}}$ and a varying precision $\sigma_R$ between 0.1 km and 1.2 km.} \label{pdfR}
\end{figure}

Mock observations are generated assuming $R_{1.4}$ has a Gaussian distribution with a mean value of $R_0=11.9~{\rm{km}}$ (supported by the LIGO/VIRGO observation for GW170817 \cite{LIGO}) with observational uncertainties ranging from $\sigma_R=(0.1\!-\!1.2)~{\rm{km}}$. For each assumed precision, an independent Bayesian inference is performed using identical priors and astrophysical constraints \cite{Li24PRD,Xavier1,Li26APJ,Xavier2}. Shown in Fig. \ref{pdfR} are the resulting posterior probability distribution functions (PDFs) of $R_{1.4}$. As expected, broader measurements lead to broader posteriors. Surprisingly, however, the posterior mean radius shifts systematically toward larger values as the measurement uncertainty increases, providing direct evidence that nonlinear filtering acts already at the observable level.
Moreover, at large uncertainties the PDFs are apparently asymmetric about the means. These findings are qualitatively consistent with the Jensen expansion of Eq. (\ref{eq:jensen}). With the approximate scaling $\rm{Pre}(2\rho_0)\propto R_{1.4}^4$, $\rm{Pre}''(R_0)\propto 12R_0^2>0$, so the inverse mapping of pressure from radius is strongly convex. Consequently, $\langle \rm{Pre}(\sigma_R)\rangle>\rm{Pre}(R_0)$ at a finite $\sigma_R$, implying that broader radius uncertainties systematically bias the inferred pressure toward larger values.

\begin{figure}[thb]
\centering
\vspace{-0.5cm}
 \resizebox{0.7\textwidth}{!}{
\includegraphics[width=1.\textwidth]{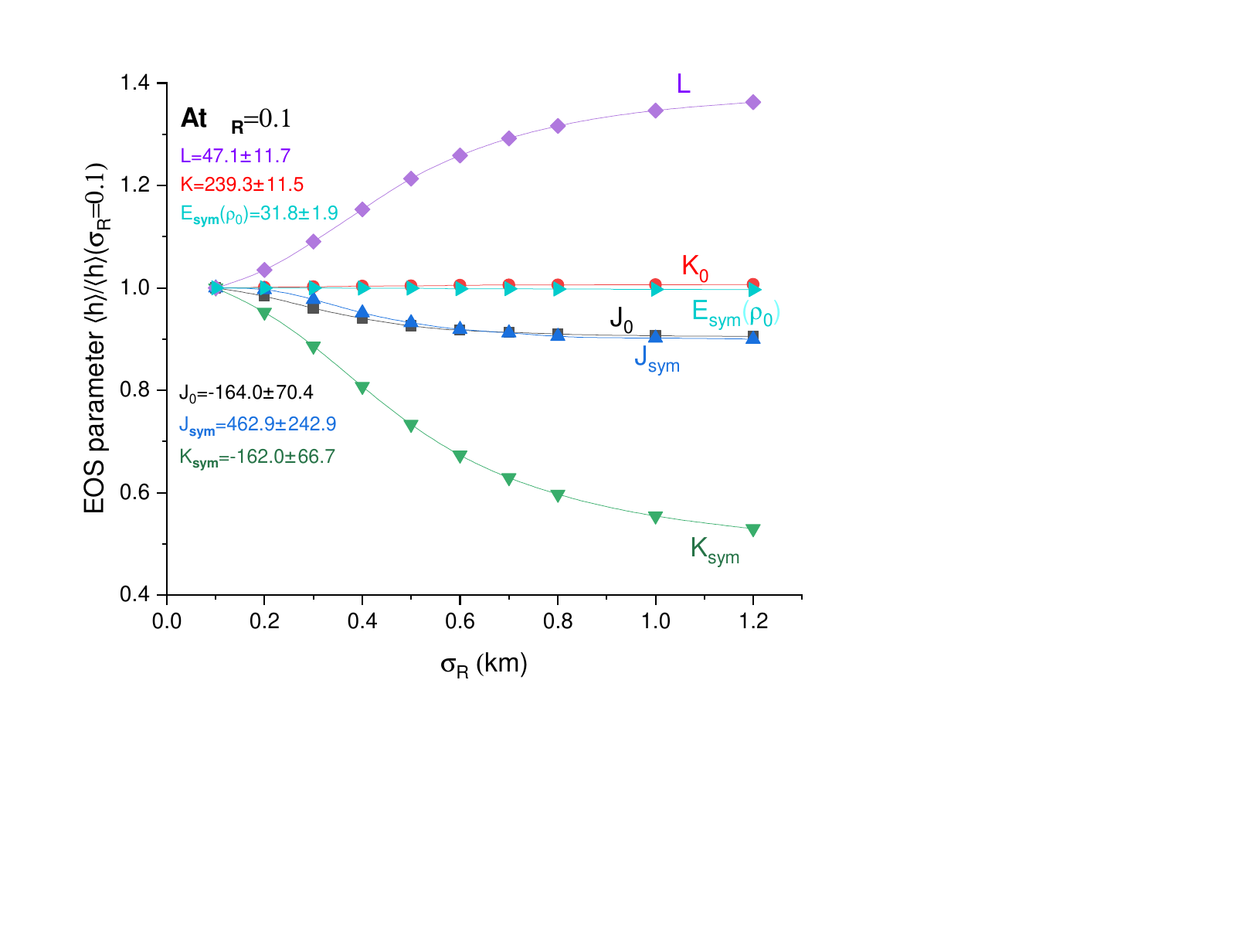}
}
\vspace{-3cm}
\caption{Posterior mean values of the indicated six EOS parameters with respect to their values at $\sigma_R=0.1$ km as functions of precision $\sigma_R.$
} \label{6means}
\end{figure}

The NS pressure is determined by all six EOS parameters through Eq. (\ref{pre}) and its value around $2\rho_0$ are known to be dominated by $L$ and $K_{\rm{sym}}$ \cite{LattimerPrakash,LCK08}. As we shall discuss, the curvature of $\rm{Pre}(2\rho_0)(R_{1.4})$ inverse mapping discussed above is primarily due to those of $L$ and $K_{\rm{sym}}$. Shown in Fig. \ref{6means} are the posterior mean values of the indicated six EOS parameters with respect to their values at $\sigma_R=0.1$ km as functions of precision $\sigma_R.$ Interestingly, their sensitivities to the radius precision are markedly different. Over the explored range of observational precisions, the inferred mean values of $L$ and $K_{\rm{sym}}$ vary most strongly, changing by approximately $37\%$ and $-47\%$, respectively. In contrast, $K_0$ and $E_{\rm{sym}}(\rho_0)$ vary only weakly by about $(1\!-\!3)\%$, while $J_0$ and $J_{\rm{sym}}$ exhibit intermediate sensitivity of approximately $10\%$. This hierarchy reflects the well-known fact that the radius of a canonical NS is controlled primarily by the symmetry energy contribution to NS pressure around $(1\!-\!2)\rho_0$, which depends most strongly on $L$ and $K_{\rm{sym}}$. \\

\textit{Nearly Universal Inverse Mappings.-} 
To construct the inverse mappings, all accepted posterior samples are first sorted according to their predictions for the canonical NS radius $R_{1.4}$ and grouped into bins of width $0.05$ km. Within each radius bin, the conditional posterior mean of an EOS parameter $h$ is calculated as
$
H(R_{1.4}) \equiv \langle h \mid R_{1.4}\rangle,
$
which estimates the conditional expectation of $h$ for a given radius $R_{1.4}$. The resulting conditional expectation functions $H(R_{1.4})$, shown in Figs.~3, 5, and 6, constitute the nearly universal EOS--radius inverse mappings. The overall posterior mean of each EOS parameter is then reconstructed by averaging these conditional means over the posterior radius distribution $P(R_{1.4},\sigma_R)$ shown in Fig. \ref{pdfR} according to the nonlinear filtering equation
\begin{equation}
\langle h\rangle(\sigma_R)
=
\int H(R_{1.4})P(R_{1.4},\sigma_R)\,dR_{1.4}.
\label{eq:nonlinear}
\end{equation}
This equation is the continuous form of the law of total expectation. Thus, two statistically distinct averaging operations are involved: a local average over posterior samples within each radius bin to determine the inverse mapping, followed by a global average over the posterior radius distribution to recover the overall posterior mean.

\begin{figure}[thb]
\centering
\vspace{-1.2cm}
 \resizebox{0.4\textwidth}{!}
 {
\includegraphics[width=0.5\textwidth]{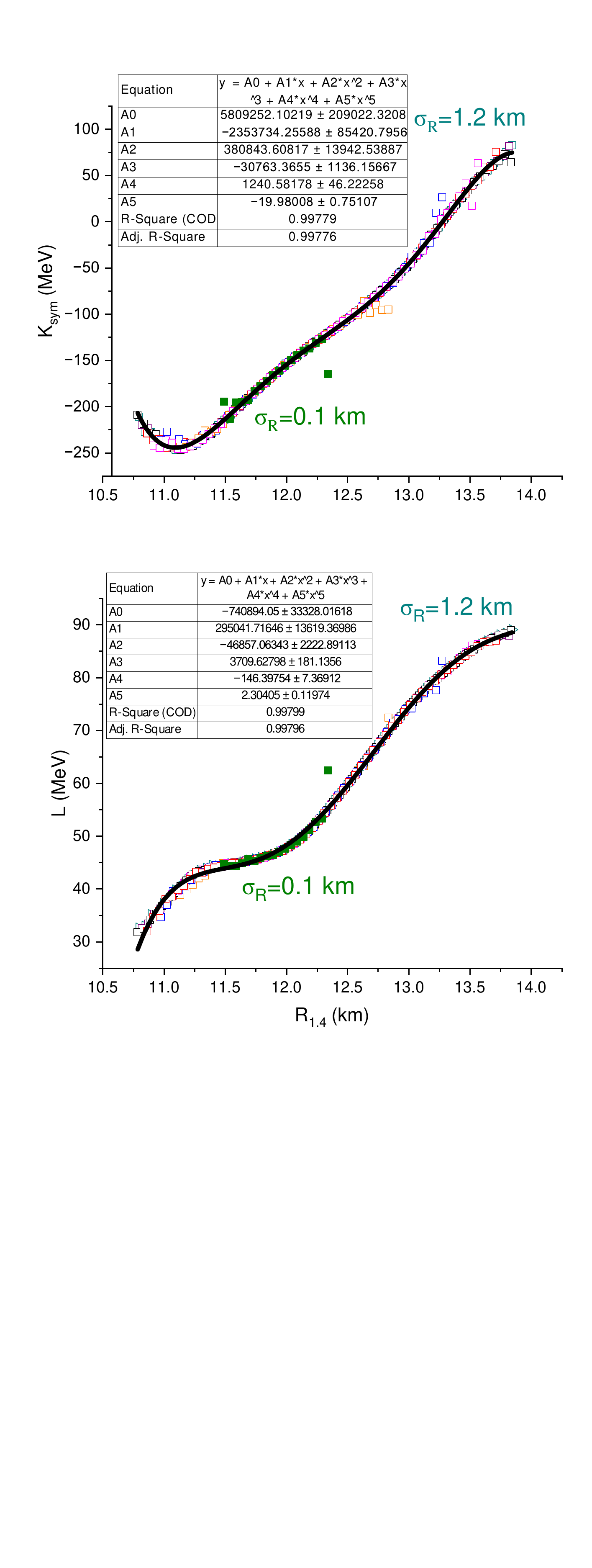}
}
\vspace{-6.50cm}
\caption{EOS-radius inverse mappings $H(R_{1.4})$ for $H=L$ and $H=K_{\rm{sym}}$, respectively, obtained using ten different observational precisions between 0.1 km and 1.2 km. A cut-off of the probability $P(R,\sigma_R)=10^{-6}$ is used in the plot. A few outliers are beyond $(3\!-\!4)\sigma_R$ values near the cut-off. The dark curves are 5th-order polynomial fits to all data points. } \label{Uscaling}
\end{figure}

Shown in Fig. \ref{Uscaling} are the EOS-radius inverse mappings for $L$ and $K_{\rm{sym}}$. These mappings obtained using ten different observational precisions between 0.1 km and 1.2 km collapse onto nearly identical curves accurately represented by fifth-order polynomials given in the insets. The corresponding plots for the four other EOS parameters are given in the End Matter. The pronounced nonlinearities visible in the high probability regions of $L(R_{1.4})$ and $K_{\rm{sym}}(R_{1.4})$ mappings therefore make them the dominant contributors to the precision-induced corrections. The existence of these nearly universal curves implies that posterior samples occupy only a narrow, low-dimensional submanifold of the six-dimensional EOS parameter space, parameterized primarily by the radius. 

An equally important observation is that the mappings are not merely approximate linear correlations. Several of them, especially $L(R_{1.4})$, $K_{\rm{sym}}(R_{1.4})$, and $J_{0}(R_{1.4})$, exhibit substantial curvature over the physically allowed radius range. Consequently, even a symmetric uncertainty in the measured radius generally leads to an asymmetric response in the inferred EOS parameters. This behavior is a direct manifestation of nonlinear filtering through the nearly universal EOS--radius manifold and provides the physical origin of the Jensen-type corrections. In particular, the curvature of the inverse mappings determines both the sign and magnitude of the precision-induced corrections, linking the topology of the EOS manifold directly to the inferred posterior means of EOS parameters.

The mappings reported here are expected to depend primarily on the constrained EOS manifold rather than on the particular meta-model parameterization adopted. The TOV equations depend only on the macroscopic pressure-energy density relation $P(\varepsilon)$ and not on its microscopic realization \cite{Li26}. Consequently, different microscopic nuclear many-body theories that generate similar EOSs and satisfy the same astrophysical constraints are expected to populate the same physically admissible region of EOS space. The inverse mappings therefore reflect primarily the geometry of the constrained EOS manifold rather than details of any particular EOS representation.

The existence of the inverse mappings is not unique to the present Bayesian framework. An analogous correlation between $L$ and $R_{1.4}$ was previously identified by Lattimer using a representative set of relativistic mean-field and Skyrme energy-density functionals \cite{Lattimer2023}. The convex behavior of the empirical $L(R_{1.4})$ relation obtained here is qualitatively consistent with these theoretical predictions. This agreement suggests that the inverse mappings uncovered in the present work reflect generic properties of the EOS--radius connection rather than artifacts of the Bayesian inference procedure or the specific EOS parameterization adopted. The mappings identified here therefore appear to emerge from the underlying physics governing NS structure rather than from the details of Bayesian inference or EOS parameterization.
\\
\begin{figure}[thb]
\centering
\vspace{-1.cm}
 \resizebox{0.5\textwidth}{!}{
\includegraphics[width=1\textwidth]{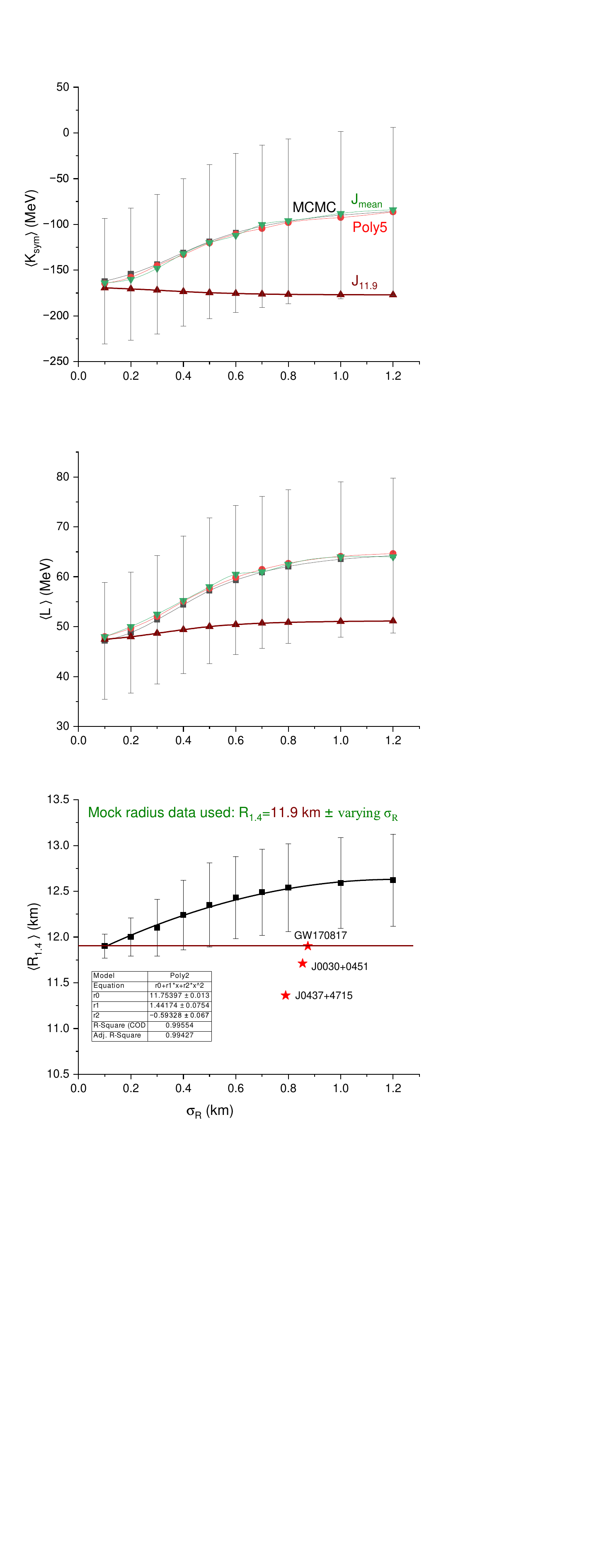}
}
\vspace{-7.0cm}
\caption{Bottom: Posterior means of $R_{1.4}$ versus $\sigma_R$ extracted from Fig. \ref{pdfR}. The current radii and $\sigma$ values of three typical canonical NSs: GW170817 \cite{LIGO}, PSR J0030+0451 \cite{451a} and PSR J0437–4715 \cite{4715a}, are shown as references. The thick black curve is a second order polynomial fit with coefficients given in the inset. Middle and upper windows: comparing the MCMC, $J_{\rm{mean}}$, Poly5 and $J_{11.9}$ values of $\langle L\rangle$ and $\langle K_{\rm{sym}}\rangle$ (see the text for their definitions) as functions of the precision $\sigma_R$, respectively. 
} \label{HJ}
\end{figure}

\textit{Quantitative Validation of Nearly Universal EOS-Radius Inverse Mappings.-} Having established the existence of the precision-independent EOS-radius inverse mappings $H(R_{1.4})$, the posterior mean of any EOS parameter $h$ can be obtained from the nonlinear filtering equation. The latter provides the exact relation connecting measurement precision and inferred EOS parameters. To test it, we reconstruct posterior means (Poly5) directly from the analytical mapping functions $H(R_{1.4})$ and the posterior radius distributions. Moreover, we perform a modified Jensen expansion around the posterior mean value $\langle R_{1.4}\rangle$ for each $\sigma_R$
\begin{equation}\label{modified}
\langle h\rangle(\sigma_R)
\simeq
h(\langle R_{1.4}\rangle(\sigma_R))
+\frac12 h''(\langle R_{1.4}\rangle(\sigma_R))\sigma_R^2,
\end{equation}
and label the result as $J_{\rm{mean}}$ in comparison with $J_{11.9}$ from Eq. (\ref{eq:jensen}) using $R_0=11.9$ km. These three values are compared with the results (label as MCMC) of full Bayesian analyses in Fig. \ref{HJ} for $L$ and $K_{\rm{sym}}$. Interestingly, both the Poly5 and $J_{\rm{mean}}$ remain remarkably accurate even for relatively broad distributions. This demonstrates that the precision dependence of inferred EOS parameters originates primarily from nonlinear filtering of the posterior radius distribution through the inverse mappings. In contrast, the local quadratic approximation of Eq.~(\ref{eq:jensen}) succeeds only for narrow radius distributions. For all six EOS parameters, the reconstructed Poly5 and $J_{\rm{mean}}$ values reproduce the full Bayesian posterior means (MCMC) with high accuracy as shown in Table \ref{Table1} of the End matter. In short, variations among currently accepted EOSs are dominated by a small number of effective degrees of freedom, with the canonical NS radius $R_{1.4}$ serving as an effective coordinate along the manifold. This explains why the inverse mappings using different observational precisions collapse onto nearly universal curves.
\\

\textit{Physical Interpretation and Implications of the Nearly Universal EOS-Radius Inverse Mappings.-} The observed universality originates from two fundamental ingredients. First, the TOV equations define a highly nonlinear mapping between the EOS and NS observables. The approximate relation $\rm{Pre}(2\rho_0)\propto R_{1.4}^4$ already implies a strong convexity. Second, physically admissible EOSs occupy a highly correlated region of parameter space because of causality, stability, nuclear-physics constraints, and NS observations. Together these effects generate a constrained EOS manifold. This interpretation naturally explains why the inverse mappings are nearly independent of observational precision. Moreover, the emergence of nearly one-dimensional inverse mappings indicates that, after imposing current laboratory, theoretical, and astrophysical constraints, the six-dimensional EOS parameter space is highly correlated. 

The strongest precision dependence occurs for the symmetry energy parameters $L$ and $K_{\rm{sym}}$ because they dominate the pressure of NS matter near $(1\!-\!2)\rho_0$, the density region to which the radius of a canonical NS is most sensitive. Consequently, the nonlinear EOS--radius mappings are most pronounced for these parameters, whereas $K_0$, $J_0$, $J_{\rm{sym}}$, and $E_{\rm{sym}}(\rho_0)$ remain comparatively insensitive to radius precision. These mappings therefore provide a quantitative framework for understanding, predicting, and correcting nonlinear-filtering effects in Bayesian EOS inference.

An important implication of the present results is that posterior means inferred from finite-precision observations should not automatically be interpreted as unbiased estimates of the underlying EOS parameters. In nonlinear inverse problems, finite observational uncertainties can generate systematic shifts whose magnitude and sign are determined by the curvature of the observable--model parameter mapping. Consequently, direct comparisons between posterior means inferred from current NS observations and microscopic nuclear many-body theories should account for nonlinear-filtering effects. In particular, the asymptotic high-precision limit may provide a more faithful estimate of the underlying EOS parameters than posterior means obtained from present observations.
\\

\textit{Summary and Conclusions.-} We have shown that Bayesian inference of NS EOSs is governed by a previously unrecognized low-dimensional constrained manifold, whose observable manifestation is a family of nearly universal EOS-radius inverse mappings. This underlying structure separates intrinsic EOS physics from measurement effects and provides a unified framework for understanding precision-dependent parameter inference. The latter originates from nonlinear filtering of the posterior radius distribution through the inverse mappings. In the narrow-distribution limit, this effect reduces naturally to a Jensen-type correction determined by the local curvature of the mapping, while for presently realistic observational uncertainties the full nonlinear-filtering relation accurately reproduces the posterior means obtained from Bayesian analyses. The framework developed here provides a practical route for correcting precision-dependent biases in EOS inference as NS observations continue to improve toward the 0.1 km era.

Consequently, posterior means inferred from current observations should not automatically be compared directly with microscopic nuclear many-body theory predictions without accounting for nonlinear-filtering corrections, whenever observational uncertainties remain significant. 
More broadly, the mechanism identified here is a generic consequence of nonlinear parameter inference from finite-precision observables and is therefore expected to arise well beyond NS physics. In particular, similar nonlinear effects are expected in analyses of neutron-skin thicknesses and collective excitations of nuclei, observables in heavy-ion collisions, and many other nonlinear inverse problems throughout nuclear physics and astrophysics. The present framework thus provides a unified perspective on precision-dependent parameter inference and opens a new avenue for connecting increasingly precise observations directly to microscopic nuclear many-body theories of supradense neutron-rich matter.
\\

\textit{Acknowledgement.-}
We thank John Holloway for helpful discussions. This work was supported in part by the U.S. Department of Energy, Office of Science, under Award No. DE-SC0013702.\\

\textit{DATA AVAILABILITY:} The data supporting the findings of this work,
including processed data and posterior samples, will be made openly available \cite{Dataset}.

\vspace{1cm}
{\bf End Matter}
\\\\
\textit{Meta-Model NS EOS.-} To explore the impact of observational precision on EOS inference while minimizing model-dependent assumptions, we employ the meta-model framework developed in Refs.~\,\cite{ZhangLi2018,ZhangLi2019,Xie2019,XieLi2020,ZhangLi2021}. Rather than adopting a specific microscopic interaction within a particular nuclear many-body theory, the meta-model parameterizes the energy per nucleon of neutron-rich matter directly in terms of widely used empirical nuclear matter parameters characterizing the density dependence of SNM and the nuclear symmetry energy. Available constraints of these parameters can be directly used in setting their priors. This meta-model can mimic the predictions of a broad class of relativistic and non-relativistic nuclear many-body approaches within a unified framework. Since the TOV equations see directly only the macroscopic EOS $P(\varepsilon)$ but not the microscopic details responsible for generating it, global NS observables are intrinsically composition degenerate and do not probe directly the underlying interactions \cite{Cai25}. Consequently, the mappings inferred from stellar observations primarily constrain the EOS $P(\varepsilon)$ itself and its derivatives rather than physical details of any specific microscopic model \cite{Li26}.
Therefore, the meta-model provides a natural intermediate representation between microscopic nuclear theories and macroscopic NS observables. These considerations suggest that the mappings should be largely independent of the specific EOS parameterization adopted here, provided the same physical constraints are imposed. The universality claimed here therefore refers to the constrained EOS manifold defined by the TOV equations and present physical constraints, rather than to a mathematical identity valid for arbitrary EOS parameterizations.

For each parameter set, the complete NS EOS is constructed by combining the meta-model parameterizations with standard constraints of thermodynamic stability, causality, charge neutrality and $\beta$-equilibrium. The resulting EOS is used to solve the TOV equations and compute NS observables. Solutions supporting the presently observed most massive NSs are accepted. More details and examples of its recent applications in Bayesian analyses can be found in Refs. \cite{Li24PRD,Xavier1,Li26APJ,Xavier2}.  
\\
\begin{table*}[t]
\caption{Comparisons of posterior means from full Bayesian analyses (MCMC), nonlinear-filtering reconstruction (Poly5) using the EOS-radius mapping, and Jensen expansion about the posterior mean radius ($J_{\rm{mean}}$).}
\scriptsize
\begin{tabular}{c|ccc|ccc|ccc|ccc|ccc|ccc}
$\sigma_R$
& \multicolumn{3}{c|}{$J_0$(MeV)}
& \multicolumn{3}{c|}{$K_0$(MeV)}
& \multicolumn{3}{c|}{$J_{\rm{sym}}$(MeV)}
& \multicolumn{3}{c|}{$K_{\rm{sym}}$(MeV)}
& \multicolumn{3}{c|}{$L$(MeV)}
& \multicolumn{3}{c}{$E_{\rm{sym},0}$(MeV)}\\
& MCMC&Poly5&$J_{\rm{mean}}$& MCMC&Poly5&$J_{\rm{mean}}$& MCMC&Poly5&$J_{\rm{mean}}$& MCMC&Poly5&$J_{\rm{mean}}$& MCMC&Poly5&$J_{\rm{mean}}$& MCMC&Poly5&$J_{\rm{mean}}$\\
\hline
0.1 & -164.0 & -164.1 & -163.2 & 239.3 & 240.3 & 240.4 & 462.9 & 470.5 & 472.0 & -162.0 & -164.6 & -164.0 & 47.1 & 48.0 & 48.0 & 31.8 & 31.0 & 31.0 \\
0.2 & -161.2 & -160.8 & -161.0 & 239.6 & 240.4 & 240.6 & 461.2 & 464.1 & 464.0 & -154.3 & -157.9 & -160.0 & 48.8 & 49.7 & 50.0 & 31.8 & 31.0 & 31.0 \\
0.3 & -157.5 & -157.3 & -157.5 & 239.8 & 240.7 & 240.8 & 452.4 & 455.1 & 452.0 & -143.5 & -145.1 & -148.0 & 51.4 & 52.0 & 52.5 & 31.8 & 30.9 & 31.0 \\
0.4 & -154.2 & -154.3 & -154.5 & 240.1 & 241.2 & 241.0 & 440.4 & 446.7 & 448.0 & -130.8 & -133.0 & -132.0 & 54.4 & 55.1 & 55.2 & 31.8 & 30.9 & 30.9 \\
0.5 & -151.8 & -152.3 & -153.0 & 240.3 & 241.4 & 241.6 & 431.3 & 437.9 & 432.0 & -118.8 & -120.2 & -120.0 & 57.2 & 57.7 & 58.0 & 31.8 & 30.8 & 30.8 \\
0.6 & -150.4 & -151.2 & -151.5 & 240.5 & 241.6 & 241.5 & 425.3 & 433.5 & 436.0 & -109.1 & -110.6 & -112.0 & 59.3 & 59.8 & 60.5 & 31.8 & 30.8 & 30.8 \\
0.7 & -149.8 & -150.4 & -152.0 & 240.6 & 241.8 & 241.5 & 422.2 & 430.6 & 428.0 & -102.0 & -104.4 & -100.0 & 60.9 & 61.5 & 61.0 & 31.8 & 30.8 & 30.8 \\
0.8 & -149.2 & -150.1 & -152.5 & 240.7 & 241.9 & 241.6 & 419.1 & 430.3 & 428.0 & -96.7 & -98.1 & -96.0 & 62.1 & 62.7 & 62.5 & 31.8 & 30.7 & 30.8 \\
1.0 & -148.6 & -149.6 & -151.0 & 240.8 & 242.1 & 242.0 & 417.6 & 428.0 & 412.0 & -89.8 & -92.4 & -88.0 & 63.5 & 64.1 & 64.0 & 31.7 & 30.7 & 30.7 \\
1.2 & -148.4 & -149.6 & -152.5 & 240.9 & 242.1 & 242.0 & 416.6 & 424.0 & 408.0 & -85.8 & -86.3 & -84.0 & 64.2 & 64.7 & 64.0 & 31.7 & 30.7 & 30.7 \\
\end{tabular}\label{Table1}
\end{table*}

\begin{figure}[thb]
\centering
\vspace{-1cm}
 \resizebox{0.4\textwidth}{!}{
\includegraphics[width=1.\textwidth]{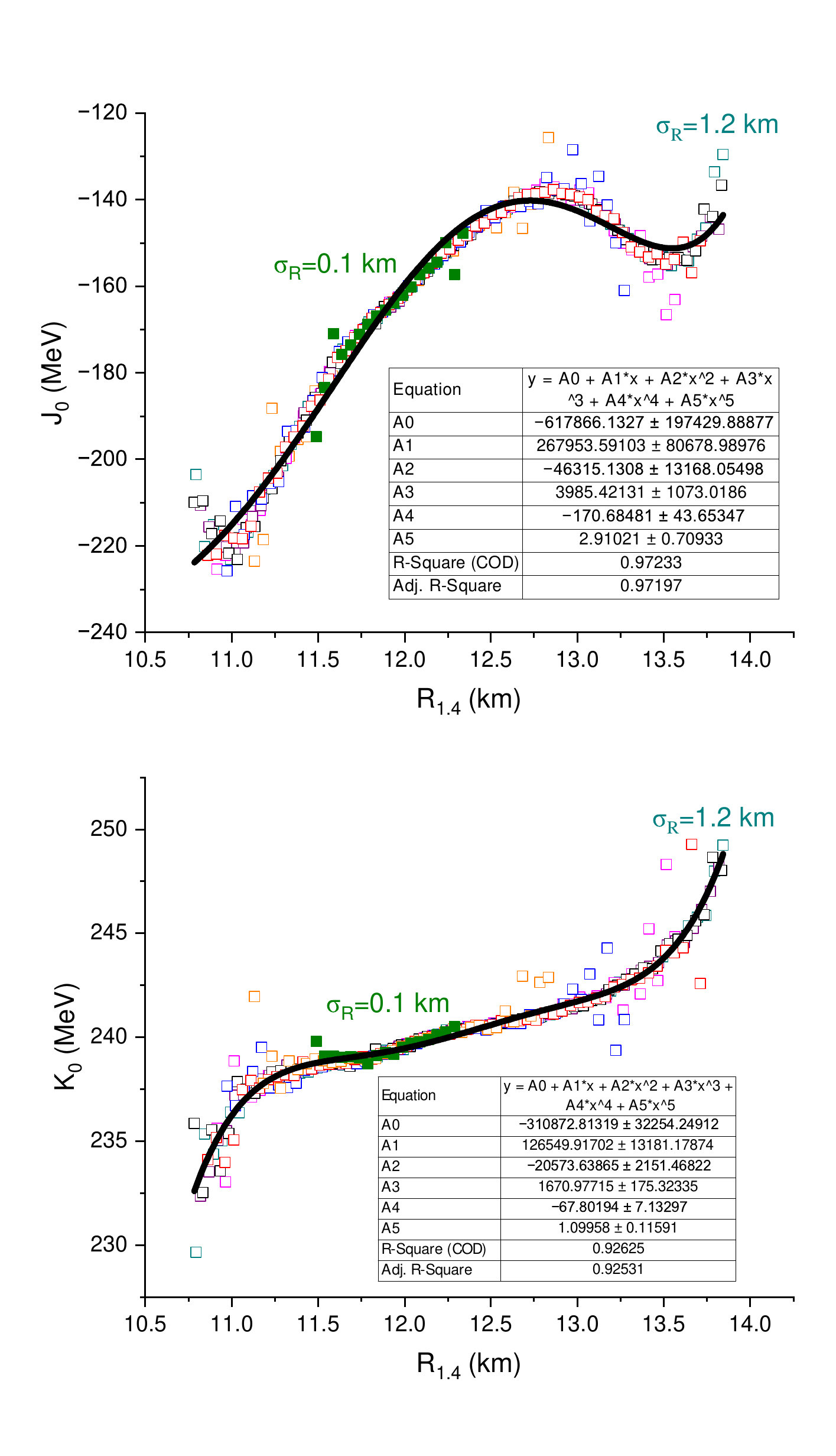}
}
\vspace{-0.8cm}
\caption{The same as in Fig. \ref{Uscaling} but for $J_0$ and $K_0$ parameters.} \label{Figures1}
\end{figure}

\begin{figure}[thb]
\centering
\vspace{-1.cm}
 \resizebox{0.4\textwidth}{!}{
\includegraphics[width=1.\textwidth]{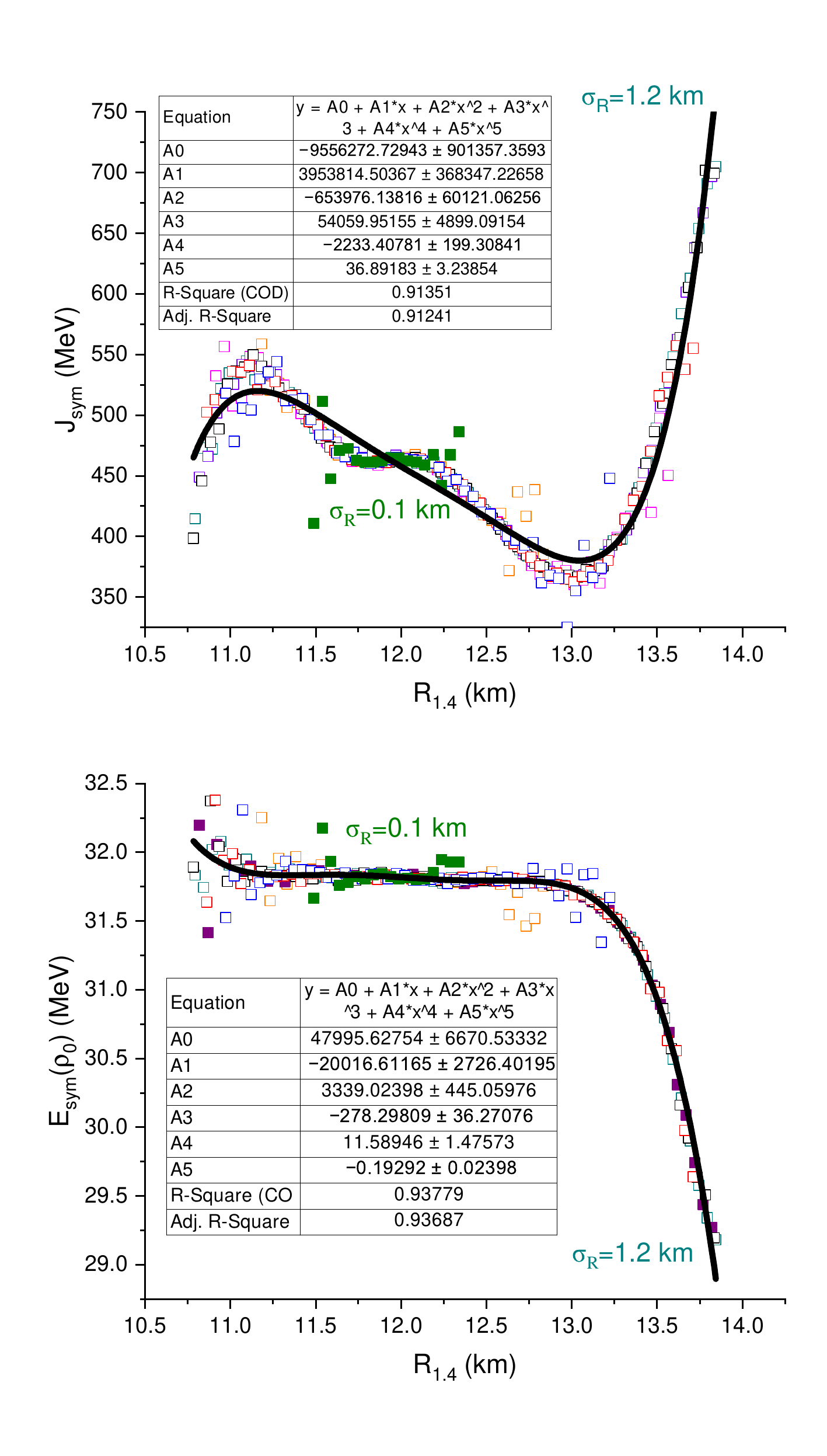}
}
\vspace{-0.8cm}
\caption{The same as in Fig. \ref{Uscaling} but for $J_{\rm{sym}}$ and $E_{\rm{sym},0}$ parameters.} \label{Figures2}
\end{figure}

\textit{Mappings of $J_0$, $K_0$, $J_{\rm{sym}}$ and $E_{\rm{sym},0}$.-} Figures~ \ref{Figures1} and \ref{Figures2} illustrate the global mappings of $J_0$, $K_0$, $J_{\rm{sym}}$ and $E_{\rm{sym},0}$ over the entire posterior radius range. Far from the posterior means, typically beyond $(3$--$4)\sigma_R$, several EOS parameters exhibit rapid but physically unimportant variations. These nontrivial structures arise primarily from strong correlations among the empirical EOS parameters required to satisfy all imposed nuclear physics and astrophysical constraints. However, within the physically relevant $(2$--$3)\sigma_R$ region where the posterior probability is appreciable, the mappings become much smoother. In particular, $J_{\rm sym}$, $K_0$, and $E_{\rm{sym}}(\rho_0)$ remain nearly constant over this interval, confirming that these quantities are only weakly constrained by the canonical NS radius, consistent with the results presented in the main text. The remaining parameters, especially $L$ and $K_{\rm{sym}}$, display pronounced but smooth nonlinear variations that are well captured by the fifth-order polynomial parameterizations. Small discrepancies between the polynomial fits and the numerical mappings, most noticeably for $J_{\rm{sym}}$, occur only in regions of extremely low posterior probability and therefore have negligible influence on the nonlinear-filtering calculations of posterior means.

\textit{Quantitative Validation of the Nearly Universal Inverse Mappings.-} To facilitate independent verification and future applications of the inverse mappings, Table \ref{Table1} compares the posterior means obtained from full Bayesian analyses (MCMC), direct nonlinear-filtering reconstruction (Poly5) using Eq. (\ref{eq:nonlinear}), and the modified Jensen approximation ($J_{\rm{mean}})$ using Eq. (\ref{modified}). 
No parameters of the polynomial fits were adjusted to reproduce the MCMC posterior means. The excellent agreement between the MCMC and nonlinear-filtering results demonstrates that the precision dependence of inferred EOS parameters is controlled almost entirely by that of the posterior radius distributions.

Once calibrated using sufficiently high-precision observations, the EOS-radius mappings may provide a computationally inexpensive surrogate for rapidly estimating EOS parameters directly from measured NS radii.

\bibliographystyle{apsrev4-2}

\end{document}